\documentclass[11pt]{article}

\usepackage{alltt}
\usepackage{graphics}
\usepackage{latexsym}
\usepackage{textcomp}
\usepackage{amssymb}
\usepackage{amsmath}
\usepackage{amsthm}
\usepackage{pst-tree}
\psset{levelsep=5ex,nodesep=3pt,radius=1ex,fillcolor=black,hatchsep=1.5pt}
\psset{tnpos=r}

\renewcommand{\leq}{\leqslant}
\renewcommand{\geq}{\geqslant}
\renewcommand{\emptyset}{\varnothing}
\newcommand{\aco}{\texttt{\symbol{'173}}}
\newcommand{\act}{\texttt{\symbol{'175}}}

\newtheorem{theorem}{Theorem}
\theoremstyle{definition}
\newtheorem{definition}{Definition}
\theoremstyle{remark}
\newtheorem{remark}{Remark}[section]

\title{On the tree-transformation power of XSLT}

\author{Wim Janssen \and
Alexandr Korlyukov$^{\text{\textdied}}$
\and
Jan Van den Bussche\thanks{Wim Janssen and Jan Van den Bussche are with the University of Hasselt, Belgium.}}

\date{}

\begin{document}
	\maketitle
	\renewcommand{\thefootnote}{}
		\footnotetext{Alexandr Korlyukov, who was with Grodno State University,
Belarus, sadly passed away shortly after we agreed to write a joint paper.}
	\renewcommand{\thefootnote}{\arabic{footnote}}

		\begin{abstract}
			XSLT is a standard rule-based programming language for expressing
			transformations of XML data.  The language is currently in transition from
			version 1.0 to 2.0.  In order to understand the computational consequences of
			this transition, we restrict XSLT to its pure tree-transformation
			capabilities.  Under this focus, we observe that XSLT~1.0 was not yet a
			computationally complete tree-transformation language: every 1.0 program can
			be implemented in exponential time.
			A crucial new feature of version~2.0,
			however, which allows node sets over temporary trees, yields completeness.  We
			provide a formal operational semantics for XSLT programs, and establish
			confluence for this semantics.		
		\end{abstract}

		\section{Introduction}
		
			XSLT is a powerful rule-based programming language, relatively widely used,
			for expressing transformations of XML data, and is developed by the W3C (World
			Wide Web Consortium) \cite{xslt,xslt2,xslt_kay}.  An XSLT program is run on
			an XML document as input, and produces another XML document as output.  (XSLT
			programs are actually called ``stylesheets'', as one of their main uses is to
			produce stylised renderings of the input data, but we will continue to call
			them programs here.)
			
			The language is actually in a transition period: the current standard,
			version~1.0, is being replaced by version~2.0.  It is important to understand
			what the new features of 2.0 really add.  In the present paper, we focus on
			the tree-transformation capabilities of XSLT\@.  Indeed, XML documents are
			essentially ordered, node-labeled trees.
			
			From the perspective of tree-transformation capabilities, the most important
			new feature is that of ``node sets over temporary trees''.  We will show that
			this feature turns XSLT into a computationally complete tree-transformation
			language.  Indeed, as we will also show, XSLT~1.0 was \emph{not} yet complete
			in this sense.  Specifically, any 1.0 program can be implemented within
			exponential time in the worst case.  Some programs actually express
			PSPACE-complete problems, because we will show that any linear-space turing
			machine can be simulated by an XSLT~1.0 program.
			
			To put our results in context, we note that the designers of XSLT will most
			probably regard the incompleteness of their language as a feature, rather than
			a defect.  Indeed, in the requirements document for 2.0, turning XSLT into a
			general-purpose programming language is explicitly stated as a ``non-goal''
			\cite{xsltreq}.  In that respect, our result on the completeness of 2.0
			exposes (albeit in a narrow sense) a failure to meet the requirements!
			
			At this point we should be a little clearer on what we mean by ``focusing on
			the tree-transformation capabilities of XSLT''\@.  As already mentioned, XML
			documents are essentially trees where the nodes are labeled by arbitrary
			strings.  We make abstraction of this string content by regarding the node
			labels as coming from some finite alphabet.  Accordingly, we strip XSLT of its
			string-manipulation functions, and restrict its arithmetic to arbitrary
			polynomial-time functions on counters, i.e., integers in the range
			$\{1,2,\dots,n\}$ with $n$ the number of nodes in the input tree.  It is,
			incidentally, quite easy to see that XSLT~1.0
			\emph{without} these restrictions
			can express all computable functions on strings (or integers).  Indeed, rules
			in XSLT can be called recursively, and we all know that arbitrary recursion
			over the strings or the integers gives us completeness.
			
			We will provide a formal operational semantics for the substantial fragment of
			XSLT discussed in this paper.  A formal semantics has not been available,
			although the W3C specifications represent a fine effort in defining it
			informally.  Of course we have tried to make our formalisation faithful to
			those specifications.  Our semantics does not impose an order on operations
			when there is no need to, and as a result the resulting transition relation is
			non-deterministic.  We establish, however, a confluence property, so that any
			two terminating runs on the same input yield the same final result.
			Confluence was not yet proven rigorously for XSLT, and can help in providing a
			formal justification for alternative processing strategies that XSLT
			implementations may follow for the sake of optimisation.

			\section{Data model}
			
			\newcommand{\X}{\mathcal{X}}
			\newcommand{\T}{\mathcal{T}}
			\newcommand{\V}{\mathcal{V}}
			\renewcommand{\t}{\mathbf{t}}
			\renewcommand{\S}{\mathbf{S}}
			\newcommand{\E}{\mathbf{E}}
			\newcommand{\n}{\mathbf{n}}
			\newcommand{\Input}{\textsf{Input}}
			\newcommand{\nodes}{\textsf{nodes}}
			\newcommand{\mixed}{\textsf{mixed}}
			\newcommand{\eval}{\mathit{eval}}
			\newcommand{\tstring}{\mathit{string}}
			\newcommand{\maketree}{\mathit{maketree}}
			\newcommand{\tta}{\texttt{a}}
			\newcommand{\ttb}{\texttt{b}}
			\newcommand{\ttc}{\texttt{c}}
			\newcommand{\ttdoc}{\texttt{doc}}
			
			\subsection{Data trees} \label{secdatatrees}
				
\begin{figure}
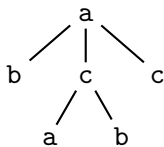

					\begin{center}
						\pstree{\TR{$\tta$}}
						{
						  \TR{$\ttb$}
						  \pstree{\TR{$\ttc$}}
						  {
						    \TR{$\tta$}
						    \TR{$\ttb$}
						  }
						  \TR{$\ttc$}
						}
					\end{center}
					\caption{A data tree.}
					\label{figdatatree}
				\end{figure}

				Let $\Sigma$ be a finite alphabet, including the special label \ttdoc.
				By a \emph{data tree} we simply mean a
				finite ordered tree, in which the nodes are labeled by elements of $\Sigma$.
				Up to isomorphism, we can describe a data tree
				$\t$ by a string $\tstring(\t)$
				over the alphabet $\Sigma$ extended with the two symbols \aco\ and \act: if
				the root of $\t$ is labeled \tta\
				and its sequence of top-level subtrees is
				$\t_1,\dots,\t_k$, then $$ \tstring(\t) = \tta \aco\tstring(\t_1)\dots
				\tstring(\t_k) \act $$
				Thus, for the data tree shown in Figure~\ref{figdatatree}, the string
				representation equals
				$$ \tta \aco \ttb \aco \act \ttc \aco \tta \aco \act \ttb \aco \act \act \ttc
				\aco \act \act. $$
				
				A \emph{data forest} is a finite sequence of data trees.  Forests arise
				naturally in XSLT, and for uniformity reasons we need to be able to
				present them as data trees.  This can easily be done as follows:
				
				\begin{definition}[maketree] \label{defmaketree}
					Let $F$ be a data forest.  Then $\maketree(F)$ is
					the data tree obtained by affixing a root node on top of $F$,
					and labeling this root node with \ttdoc.\footnote{The root node added by
					$\maketree$ models what is called the ``document root'' in the XPath data
					model \cite{xquerypath_datamodel}, although we do not model it entirely
					faithfully, as we do not formally distinguish ``document nodes'' from
					``element nodes''.  This is only for simplicity; it is no problem to
					incorporate this distinction in our formalism, and our technical results do
					not depend on our simplification.}
				\end{definition}
				
			\subsection{Stores and values}
				
				Let $\T$ be a supply of \emph{tree variables},
				including the special tree variable \Input.  We define:
				
				\begin{definition}
					A \emph{store} is a finite set $\S$ of pairs of the form
					$(x,\t)$, where $x \in \T$ and $\t$ is a data tree, such that
					(1)~\Input\ occurs in $\S$;
					(2)~no tree variable occurs twice in $\S$; and
					(3)~all data trees occurring in $\S$ have disjoint sets of nodes.
					
					The tree assigned to \Input\ is called the \emph{input tree}; the
					other trees are called the \emph{temporary trees}.
				\end{definition}
				
				\begin{definition}
					A \emph{value over $\S$} is a finite sequence consisting
					of nodes from trees in $\S$, and counters
					over $\S$.  Here, a \emph{counter over $\S$} is an integer in the range
					$\{1,2,\dots,n\}$, where $n$ is the total number of nodes in $\S$.
				\end{definition}
				
				Values as defined above formalise the kind of values that can be returned by
				XPath expressions.  XPath \cite{xpath,xpath2} is a language that is used as a
				sublanguage in XSLT for the purpose of selecting nodes from trees.  But XPath
				expressions can also return numbers, which is useful as an aid in making
				node selections (e.g., the $i$-th child of a node, or the $i$-th
				node of the tree in preorder).  We limit these numbers to counters, in order
				to concentrate on pure tree transformations.

		\section{XPath abstraction}
			
			\newcommand{\rootexp}{\texttt{/*}}
			\newcommand{\childexp}{\texttt{child::*}}
			
			Since the language XPath is already well understood
			\cite{wadler_xpath,xpath_nutshell,xquery_formal,hidders_light}, and its study
			in itself is not our focus, we will work with an abstraction of XPath, which
			we denote by $\X$.  For our purposes it will suffice to divide the
			$\X$-expressions in only two different types, which we denote by \nodes\ and
			\mixed.  A value is of type \nodes\ if it consists exclusively of
			nodes; otherwise it is of type \mixed.
			
			In order to define the semantics of $\X$, we need some definitions, which
			reflect those from the XPath specification.
			Let $\V$ be a supply of \emph{value variables}, disjoint from $\T$.
			
			\begin{definition}
				An \emph{environment over $\S$}
				is a finite set $\E$ of pairs of the form
				$(x,v)$, where $x \in \V$ and $v$ is a value over $\S$, such that no value
				variable occurs twice in $\E$.
			\end{definition}
			
			\begin{definition}
				A \emph{context triple over $\S$} is a triple $(z,i,k)$ where $z$ is a node
				from $\S$ or a counter over $\S$,
				and $i$ and $k$ are counters over $\S$ such that $i \leq k$.
				We call $z$ the \emph{context item}, $i$ the \emph{context position}, and $k$
				the \emph{context size}.
			\end{definition}
			
			\begin{definition}
				A \emph{context} is a triple $(\S,\E,c)$ where $\S$ is a store, $\E$ is an
				environment over $\S$, and $c$ is a context triple over $\S$.
			\end{definition}
			
			If we denote the universe of all possible contexts by \textit{Contexts}, the
			semantics of $\X$ is now given by a partial function $\eval$ on $\X \times
			\mathit{Contexts}$, such that whenever defined, $\eval(e,C)$ is a value over
			$C$'s store, and this value has the same type as $e$.
			
			\begin{remark} \label{xmlschema}
				A static type
				system, based on XML Schema \cite{xmlschema,essencexml}, can be put on
				contexts to ensure definedness of expressions \cite{xquery_formal}, but we
				omit that as safety is not the focus of the present paper.
				\qed
			\end{remark}
			
			In general we do not assume much from $\X$, except for the
			availability of the
			following basic expressions, also present in real XPath:
			\begin{itemize}
				\item
					An expression `\rootexp', such that $\eval(\rootexp,C)$ equals
					the root node of the input tree in $C$'s store.
				\item
					An expression `\childexp', such that $\eval(\childexp,\allowbreak C)$
					is defined whenever $C$'s context item is a node $\n$, and then
					equals the list of children of $\n$.
			\end{itemize}

		\section{Syntax} \label{secsyntax}
			
			In this section, we define the syntax of a sizeable fragment of XSLT~2.0.  The
			reader familiar with XSLT will notice that we have simplified and cleaned up
			the language in a few places.  These modifications are only for the sake of
			simplicity of exposition, and our technical results do not depend on them.  We
			discuss our deviations from the real language further in
			Section~\ref{deviations}.
			
			Also, the concrete syntax of real
			XSLT is XML-based and rather unwieldy.  For the sake of presentation, we
			therefore give a syntax of our own, which is non-XML, but otherwise follows the
			same lines as the real syntax.
			
			The grammar is shown in Figure~\ref{figgrammar}.  The only typing condition we
			need is that in an apply-statement or in a vcopy-statement, \textit{expr} must
			be of type \nodes.  Also, no two different rules can have the same
			\textit{name}, and the \textit{name} in a call-statement must be the
			\textit{name} of some rule.
						
			\begin{figure}
				\begin{tabbing}
					Program $\to$ Rule* \\
					Rule $\to$ \=\textsf{template} \textit{name} \textsf{match} \textit{expr} (\textsf{mode} \textit{name})? 
																\aco\ Template \act \\
					Template $\to$ Statement* \\
					Statement $\to$ \=$|$ \=\textsf{cons}
                                        \textit{label} \aco\ Template
                                        \act\kill
					Statement $\to$ \>\> \textsf{cons} \textit{label} \aco\ Template \act\\
						\> $|$ \textsf{apply} \textit{expr} (\textsf{mode} \textit{name})?\\
						\> $|$ \textsf{call} \textit{name}\\
						\> $|$ \textsf{foreach} \textit{expr} \aco\ Template \act\\
						\> $|$ \textsf{val} \textit{value\_variable} \textit{expr}\\
						\> $|$ \textsf{tree} \textit{tree\_variable} \aco\ Template \act\\
						\> $|$ \textsf{vcopy} \textit{expr}\\
						\> $|$ \textsf{tcopy} \textit{tree\_variable}\\
						\> $|$ \textsf{if} \textit{expr} \aco\ Template \act\ \textsf{else} \aco\ Template \act 
					
				\end{tabbing}
				\caption{Our syntax.
				The terminal symbol \textit{expr} stands for an
				$\X$-expression; \textit{label} stands for an element of our alphabet
				$\Sigma$; \textit{value\_variable} and \textit{tree\_variable} stand for
				elements of $\V$ and $\T$, respectively; and \textit{name} is
				self-explanatory.  As usual we use * to denote repetition, ? to denote
				optionality, and use (~and ) for lexical grouping.}
				\label{figgrammar}
			\end{figure}
			
			We will often identify a template $M$ with its \emph{syntax tree}.  This tree
			consists of all occurrences of statements in $M$ and represents how they
			follow each other and how they are nested in each other; we omit the formal
			definition.  Observe that only cons-, foreach-, tree-, and if-statements can
			have children.
			Note also that, since a
			template is a sequence of statements, the syntax ``tree'' is actually a
			forest, i.e., a sequence of trees, but we will still call it a tree.
			
			Variable definitions happen through val- and tree-statements.
			We will need the notion of a statement being in the scope of some variable
			definition; this is defined in the standard way as follows.
			
\begin{figure}
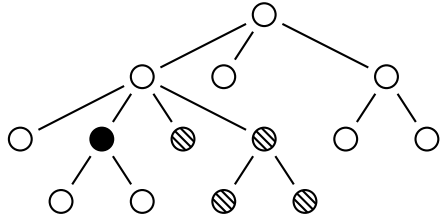

				\begin{center}
					\pstree{\TC}
					{
					  \pstree{\TC}
					  {
					    \TC
					    \pstree{\TC[fillstyle=solid]}
					    {
					      \TC
					      \TC
					    }
					    \TC[fillstyle=vlines]
					    \pstree{\TC[fillstyle=vlines]}
					    {
					      \TC[fillstyle=vlines]
					      \TC[fillstyle=vlines]
					    }
					  }
					  \TC
					  \pstree{\TC}
					  {
					    \TC
					    \TC
					  }
					}
				\end{center}
				\caption{Depiction of a syntax tree.  The nodes in the scope of the black node
				are those that are striped.}
				\label{figscope}
			\end{figure}

			\begin{definition}
				Let $M$ be a template, and let $S_1$ and $S_2$ be two
				statements occurring in $M$.  We say that $S_2$ is \emph{in the scope of}
				$S_1$ if $S_2$ is a right sibling of $S_1$
				in the syntax tree of $M$, or a descendant of such a right sibling.
				An illustration is in Figure~\ref{figscope}.
			\end{definition}
			
			One final definition:
			
			\begin{definition}
				Template $M'$ is called a \emph{subtemplate} of template $M$ if $M'$ consists
				of a sequence of consecutive sibling statements occurring in  $M$.
			\end{definition}

\section{Operational semantics} \label{secsemantics}

\newcommand{\context}{\textit{context}}
\newcommand{\confrepl}[2]{(#1\gets #2)}

Fix a program $P$ and a data tree $\t$.  We will describe the semantics of $P$
on input $\t$ as a rewrite relation $\Rightarrow$ among configurations.

\begin{definition} \label{defconfig}
A \emph{configuration} consists of a template $M$
together with a partial function that assigns a context to some of
the statements of $M$ (more precisely, the nodes of its syntax tree).
The statements that have a context are called \emph{active}; we require that
the descendants of an inactive node are inactive too.
Cons-statements are never active.

We use the following notation concerning configurations:
\begin{itemize}
\item
$S \lhd \gamma$ denotes that $S$ is a statement occurring in the template of
configuration $\gamma$.
\item
If $S \lhd \gamma$, then
$\gamma(S) = C$ denotes that $S$ is active in $\gamma$, having context $C$.
\item
If $M$ is a subtemplate of a configuration $\gamma$, then $M$
itself can be taken as a configuration by inheriting all the context
assignments done by $\gamma$.  We call such a configuration a
\emph{subconfiguration}.
\item
If $M$ is a subconfiguration of $\gamma$, and $\gamma'$ is another
configuration, then
$\gamma\confrepl{M}{\gamma'}$
denotes the configuration obtained from $\gamma$ by
replacing $M$ by $\gamma'$.
\end{itemize}
\end{definition}

The initial configuration is defined as follows.

\begin{definition}
\begin{enumerate}
\item
The \emph{initial context} equals
$$ \bigl ( 
\{(\Input,\t)\},\,\emptyset,\,(\mathbf{r},1,1)
\bigr ) $$
where $\mathbf{r}$ is the root of $\t$.
\item
The \emph{initial template} equals the single statement `\textsf{apply}
\rootexp'.
\item
The \emph{initial configuration} consists of the initial template, whose
single statement is assigned the initial context.
\end{enumerate}
\end{definition}

The goal will be to rewrite the initial configuration into a \emph{terminal
template}; this is a configuration consisting exclusively of cons-statements.
Observe that terminal templates can be viewed as data forests; indeed, simply
by removing the \textsf{cons}'s from a terminal template, we obtain the string
representation of a data forest.  

For the rewrite relation $\Rightarrow$ we are going to define, terminal
configurations will be normal forms, i.e., cannot be rewritten further.  If,
for two configurations $\gamma_0$ and $\gamma_1$, we have $\gamma_0
\Rightarrow \cdots \Rightarrow \gamma_1$ and $\gamma_1$ is a normal form, we
denote that by $\gamma_0 \Rightarrow^! \gamma_1$.  The relation $\Rightarrow$
will be defined in such a way that if $\gamma_0$ is the initial configuration
and $\gamma_0 \Rightarrow^! \gamma_1$, then $\gamma_1$ will be terminal.
Moreover, we will prove in Theorem~\ref{confluence} that each configuration
$\gamma_0$ has at most one such normal form $\gamma_1$.  We thus define:

\begin{definition} \label{deffinal}
Given $P$ and $\t$, let $\gamma_0$ be the initial configuration and let
$\gamma_0 \Rightarrow^! \gamma_1$.  Then the \emph{final result
tree of applying $P$ to $\t$} is defined to be $\maketree(\gamma_1)$.
\end{definition}

In the above definition, we can indeed apply $\maketree$, defined on data
forests (Definition~\ref{defmaketree}),
to $\gamma_1$, since $\gamma_1$ is terminal and we just observed that
terminal templates describe forests.
Note that the final result tree is only determined up to isomorphism.

\subsection{If-statements}

\newcommand{\transif}{\stackrel{\mathsf{if}}{\Rightarrow}}
\newcommand{\normif}{\mathrel{\transif{}^!}}

If-statements are the only ones that generate control flow, so we treat
them by a separate rewrite relation $\transif$, defined
by the semantic rules shown in Figure~\ref{figif}.

\begin{figure}
\begin{itemize}
\item
$
\begin{array}[t]{l}
S = \text{\textsf{if} $e$ \aco\ $M_{\rm true}$ \act\
\textsf{else} \aco\ $M_{\rm false}$ \act} \lhd \gamma \\
\gamma(S)=C \\
\eval(e,C) \neq \emptyset \\
\hline
\gamma \transif \gamma\confrepl{S}{M_{\rm true}}
\end{array}
$
\medskip
\item
$
\begin{array}[t]{l}
S = \text{\textsf{if} $e$ \aco\ $M_{\rm true}$ \act\
\textsf{else} \aco\ $M_{\rm false}$ \act} \lhd \gamma \\
\gamma(S)=C \\
\eval(e,C) = \emptyset \\
\hline
\gamma \transif \gamma\confrepl{S}{M_{\rm false}}
\end{array}
$
\end{itemize}
\caption{Semantics of if-statements; $\emptyset$ denotes the empty sequence.}
\label{figif}
\end{figure}

It is not difficult to show that $\transif$ is terminating and
locally confluent, whence confluent,
so that every configuration has a unique normal form w.r.t.\
$\transif$ \cite{terese}.  This normal form no longer contains any active
if-statements.  (Quite obviously, the most efficient way to get to this normal 
form is to work out the if-statements top-down.)
We write $\gamma \normif \gamma'$ to denote that $\gamma'$ is the
normal form of $\gamma$ w.r.t.\ $\transif$.

\begin{remark}
Our main rewrite relation $\Rightarrow$ is not terminating in general.
The reason why we treat if-statements separately is to avoid 
nonsensical rewritings such as
where we execute a non-terminating statement
in the else-branch of an if-statement whose test 
evaluates to true.
\end{remark}

\subsection{Apply-, call-, and foreach-statements}

\newcommand{\init}{\mathit{init}}

For the semantics of apply-statements, we need the following definitions.

\begin{definition}[ruletoapply] \label{defruletoapply}
Let $C$ be a context, let
$\n$ be a node, and let $m$ be a name.
Then $\mathit{ruletoapply}(C,\n)$ (respectively,
$\mathit{ruletoapply}(C,\n,m)$)
equals the template belonging to the
first rule in $P$ (respectively, with \textsf{mode} name equal to $m$)
whose \textit{expr} satisfies 
$\n \in \eval(\textit{expr},C)$.

If no such rule exists, both $\mathit{ruletoapply}(C,\n)$
and $\mathit{ruletoapply}(C,\n,m)$ default to the
single-statement template `\textsf{apply} \childexp'.
\end{definition}

\begin{definition}[init] \label{definit}
Let $M$ be a template, and let $C$ be a context.  Then $\init(M,C)$
equals the configuration obtained from $M$ by assigning context $C$ to every
statement in $M$, except for all statements in the scope of any variable
definition, and all statements that are below a foreach-statement;
all those statements remain inactive.
\end{definition}

We are now ready for the semantic rule for apply-statements, shown in
Figure~\ref{figapply}.  We omit the rule for an apply-statement with a
\textsf{mode $m$}: the only difference with the rule shown
is that we use $\mathit{ruletoapply}(\n_i,C,m)$.

\begin{figure}
\begin{itemize}
\item
$
\begin{array}[t]{l}
S = \text{\textsf{apply $e$}} \lhd \gamma \\
\gamma(S) = C = (\S,\E,c) \\
\eval(e,C) = (\n_1,\dots,\n_k) \\
\mathit{ruletoapply}(\n_i,C) = M_i \quad \text{for $i=1,\dots,k$} \\
\init(M_i,(\S,\E,(\n_i,i,k))) = \gamma_i \quad \text{for $i=1,\dots,k$} \\
\gamma\confrepl{S}{\gamma_1\dots\gamma_k} \normif \gamma' \\
\hline
\gamma \Rightarrow \gamma'
\end{array}
$
\medskip
\item
$
\begin{array}[t]{l}
S = \text{\textsf{foreach} $e$ \aco\ $M$ \act} \lhd \gamma \\
\gamma(S) = C = (\S,\E,c) \\
\eval(e,C) = (z_1,\dots,z_k) \\
\init(M,(\S,\E,(z_i,i,k))) = \gamma_i \quad \text{for $i=1,\dots,k$} \\
\gamma\confrepl{S}{\gamma_1\dots\gamma_k} \normif \gamma' \\
\hline
\gamma \Rightarrow \gamma'
\end{array}
$
\medskip
\item
$
\begin{array}[t]{l}
S = \text{\textsf{call} \textit{name}} \lhd \gamma \\
\gamma(S) = C \\
\mathit{rulewithname}(\mathit{name}) = M \\
\init(M,C) = \gamma_1 \\
\gamma\confrepl{S}{\gamma_1} \normif \gamma' \\
\hline
\gamma \Rightarrow \gamma'
\end{array}
$
\end{itemize}
\caption{Semantics of apply-, call-, and foreach-statements.}
\label{figapply}
\end{figure}

The semantic rule for foreach-statements is very similar to that for
apply-statements, and is also shown in Figure~\ref{figapply}.

For call-statements, we need the following definition.

\begin{definition}[rulewithname] \label{defrulewithname}
	For any \textit{name}, let $\mathit{rulewithname}(\mathit{name})$ denote the
template of the rule in $P$ with that \textit{name}.
\end{definition}

The semantic rule for a call-statement is then again shown in Figure~\ref{figapply}.

\subsection{Variable definitions}

\newcommand{\updateset}{\mathit{updateset}}
\newcommand{\contupd}[2]{(#1\colon #2)}

For a context $C=(\S,\E,c)$, a value variable $x$, a value $v$, a tree
variable $y$, and a data tree $\t$, we denote by
\begin{itemize}
\item
$C\contupd{x}{v}$
the context obtained from $C$ by updating $\E$ with the pair $(x,v)$;
and by
\item
$C\contupd{y}{\t}$
the context obtained from $C$ by updating $\S$ with the pair
$(y,\t)$.
\end{itemize}

We also define:

\begin{definition}[updateset] \label{defupdateset}
Let $\gamma$ be a configuration and let $S \lhd \gamma$.
Let $M$ be the template underlying $\gamma$.
Let $S_1,\dots,S_k$ be the right siblings of $S$ in $M$, in that
order.  Let $j$ be the smallest index for which $S_j$ is active in $\gamma$;
if all the $S_i$ are inactive, put $j=k+1$.  Then the template $S_1 \dots
S_{j-1}$ is denoted by $\updateset(\gamma,S)$.  If $j=1$ then this is the
empty template.
\end{definition}

We are now ready for the semantic rules for variable definitions,
shown in Figure~\ref{figvar}.

\begin{figure}
\begin{itemize}
\item
$\begin{array}[t]{l}
S = \text{\textsf{val} $x$ $e$} \lhd \gamma \\
\gamma(S) =  C \\
C\contupd{x}{\eval(e,C)} = C' \\
\updateset(\gamma,S) = M \\
\init(M,C') = \gamma_1 \\
\gamma\confrepl{SM}{\gamma_1} \normif \gamma' \\
\hline
\gamma \Rightarrow \gamma'
\end{array}$
\medskip
\item
$\begin{array}[t]{l}
S = \text{\textsf{tree} $y$ \aco\ $M$ \act} \lhd \gamma \\
\text{$M$ is terminal} \\
\gamma(S) =  C \\
C\contupd{y}{\maketree(M)} = C' \\
\updateset(\gamma,S)=M' \\
\init(M',C') = \gamma_3 \\
\gamma\confrepl{SM'}{\gamma_3} \normif \gamma' \\
\hline
\gamma \Rightarrow \gamma'
\end{array}$
\end{itemize}
\caption{Semantics of variable definitions.}
\label{figvar}
\end{figure}

\subsection{Copy-statements}

\newcommand{\forest}{\mathit{forest}}
\newcommand{\ttemp}{\mathit{ttemp}}
\newcommand{\choproot}{\mathit{choproot}}

The following definitions are illustrated in Figure~\ref{figforest}.

\begin{definition}[forest] \label{defforest}
Let $\S$ be a store, and let $(\n_1,\dots,\n_k)$ be a sequence of nodes from
$\S$.  For $i=1,\dots,k$, let $\t_i$ be the data subtree rooted at $\n_i$.
Then $\forest((\n_1,\dots,\n_k),\S)$ equals the data forest
$(\t_1,\dots,\t_n)$.
\end{definition}

\begin{definition}[ttemp] \label{defttemp}
Let $F$ be a data forest.  Then $\ttemp(F)$ equals the
terminal template describing $F$.
\end{definition}

\begin{figure}
\begin{center}
\pstree{\TR{$\tta$}}
{
  \TR{$\ttb$}~{$\n_1$}
  \pstree{\TR{$\ttc$}~{$\n_2$}}
  {
    \TR{$\tta$}~{$\n_3$}
    \TR{$\ttb$}
  }
  \TR{$\ttc$}
}
\qquad
\pstree{\TR{$\ttc$}~{$\n_4$}}
{
  \TR{$\tta$}
  \TR{$\ttb$}
}
\end{center}

\begin{tabbing}
	$ \ttemp \bigl ( \forest((\n_4,\n_1,\n_2,\n_3,\n_1),\S) \bigr ) = {}$\=	
	\textsf{cons $\ttc$ \aco\ cons $\tta$ \aco \act\ cons $\ttb$ \aco \act\ \act}\\
	\>\textsf{cons $\ttb$ \aco \act}\\
	\>\textsf{cons $\ttc$ \aco\ cons $\tta$ \aco \act\ cons $\ttb$ \aco \act\ \act}\\
	\>\textsf{cons $\tta$ \aco \act}\\
	\>\textsf{cons $\ttb$ \aco \act}
\end{tabbing}
\caption{Illustration of Definitions \ref{defforest} and \ref{defttemp}.}
\label{figforest}
\end{figure}

We also need:

\begin{definition}[choproot]
Let $\t$ be a data tree with top-level subtrees $\t_1,\dots,\t_k$, in that
order.  Then $\choproot(\t)$ equals the data forest $(\t_1,\dots,\t_k)$.
\end{definition}

The semantic rules for copy-statements are now shown in Figure~\ref{figcopy}.

\begin{figure}
\begin{itemize}
\item
$
\begin{array}[t]{l}
S = \text{\textsf{vcopy} $e$} \lhd \gamma \\
\gamma(S) = C = (\S,\E,c) \\
\eval(e,C) = (\n_1,\dots,\n_k) \\
\ttemp \bigl ( \forest((\n_1,\dots,\n_k),\S) \bigr ) = M \\
\hline
\gamma \Rightarrow \gamma\confrepl{S}{M}
\end{array}
$
\medskip
\item
$
\begin{array}[t]{l}
S = \text{\textsf{tcopy} $y$} \lhd \gamma \\
\gamma(S) = (\S,\E,c) \\
(y,\t) \in \S \\
\ttemp(\choproot(\t)) = M \\
\hline
\gamma \Rightarrow \gamma\confrepl{S}{M}
\end{array}
$
\end{itemize}
\caption{Semantics of copy-statements.}
\label{figcopy}
\end{figure}

\subsection{Discussion} \label{deviations}

The final result of applying $P$ to $\t$ (Definition~\ref{deffinal}) may be
undefined for two very different reasons.  The first, fundamental, reason is
that the rewriting may be nonterminating.  The second reason is that the
rewriting may abort because the evaluation of an $\X$-expression is undefined,
or the tree variable in a tcopy-statement is not defined in the store.  This
second reason can easily be avoided by a type system on $\X$, as already
mentioned in Remark~\ref{xmlschema}, together with scoping rules to keep track
of which variables are visible in the XSLT program and which variables are
used in the $\X$-expressions.  Such scoping rules are entirely standard, and
indeed are implemented in the XSLT processor SAXON \cite{saxon}.

In the same vein, we have simplified the parameter passing mechanism of XSLT,
and have omitted the feature of global variables.  On the other hand, our
mechanism for choosing the rule to apply (Definition~\ref{defruletoapply}) is
more powerful than the one provided by XSLT, as ours is context-dependent.
It is actually easier to define that way.  As already
mentioned at the beginning of Section~\ref{secsyntax}, none of our technical
results depend on the modifications we have made.

Finally, we note that the XSLT processor SAXON evaluates variable definitions
lazily, whereas we simply evaluate them eagerly.  Again, lazy evaluation could
have been easily incorporated in our formalism.  Some programs may terminate
on some inputs lazily, while they do not terminate eagerly, but for programs
that use all the variables they define there is no difference.

\subsection{Confluence}

\newcommand{\R}{\mathcal{R}}

Recall that we call a rewrite relation \emph{confluent} if, whenever we can
rewrite a configuration $\gamma_1$ to $\gamma_2$ as well as to $\gamma_3$,
then there exists $\gamma_4$ such that we can further rewrite both $\gamma_2$
and $\gamma_3$ into $\gamma_4$.  Confluence guarantees that all terminating
runs from a common configuration also end in a common configuration
\cite{terese}.  Since, for our rewrite relation $\Rightarrow$,
either all runs on some input are nonterminating, or none is, the
following theorem implies that the same final result of a program $P$ on an
input $\t$, if defined at all, will be obtained regardless of the order in
which we process active statements.

\begin{theorem} \label{confluence}
Our rewrite relation $\Rightarrow$ is confluent.
\end{theorem}

\begin{proof}

The proof is a very easy application of
a basic theorem of Rosen about subtree replacement systems
\cite{rosen_cr}.  A subtree replacement system $\R$ is a (typically infinite)
set of pairs of the form $\phi \to \psi$, where $\phi$ and $\psi$ are
descriptions up to isomorphism of ordered, node-labeled trees, where the node
labels come from some (again typically infinite) set $V$.  Let us refer to
such trees as $V$-trees.  Such a system $\R$ naturally induces a rewrite
system $\Rightarrow_{\R}$ on $V$-trees: we have $\t \Rightarrow_{\R} \t'$ if
there exists a node $\n$ of $\t$ and a pair $\phi \to \psi$ in $\R$ such that
the subtree $\t/\n$ is isomorphic to $\phi$, and $\t' =
\t\confrepl{\n}{\psi}$.  Here, we use the notation $\t/\n$ for the subtree of
$\t$ rooted at $\n$, and the notation $\t\confrepl{\n}{\psi}$ for the tree
obtained from $\t$ by replacing $\t/\n$ by a fresh copy of $\psi$.  Rosen's
theorem states that if $\R$ is ``unequivocal'' and ``closed'', then
$\Rightarrow_{\R}$ is confluent.

``Unequivocal'' means that for each $\phi$ there is at most one $\psi$ such
that $\phi\to \psi$ is in $\R$.  The definition of $\R$ being ``closed'' is a
bit more complicated.  To state it, we need the notion of a \emph{residue map}
from $\phi$ to $\psi$.  This is a mapping $r$
from the nonroot nodes of $\phi$ to
sets of nonroot nodes of $\psi$, such that for $m \in r(n)$ the subtrees
$\phi/n$ and $\psi/m$ are isomorphic.  Moreover, if $n_1$ and $n_2$ are
independent (no descendants of each other), then all nodes in $r(n_1)$ must
also be independent of all nodes in $r(n_2)$.

Now $\R$ being closed means that we can assign a residue map $r[\phi,\psi]$
to every $\phi\to\psi$ in $\R$ in such a way that for any $\phi_0 \to \psi_0$
in $\R$, and any node $n$ of $\phi_0$, if there exists a pair $\phi_0/n \to
\psi$ in $\R$, then the pair $\phi_0\confrepl{n}{\psi} \to
\psi_0\confrepl{r[\phi_0,\psi_0](n)}{\psi}$ is also in $\R$.  Denoting the
latter pair by $\phi_1 \to \psi_1$, we must moreover have for each node $p$
of $\phi_0$ that is independent of $n$, that $r[\phi_1,\psi_1](p) =
r[\phi_0,\psi_0](p)$.

To apply Rosen's theorem, we view configurations (Definition~\ref{defconfig})
as $V$-trees, where $V = \mathit{Statements} \cup (\mathit{Statements} \times
\mathit{Contexts})$.  Here, $\mathit{Statements}$ is the set of all possible
syntactic forms of statements.  So, given a configuration, we take the syntax
tree of the underlying template, and label every inactive node by its
corresponding statement, and every active node by its corresponding statement
and its context in the configuration.  (Since templates are sequences, we
actually get $V$-forests rather than $V$-trees, but that is a minor fuss.)

Now consider the subtree replacement system $\R$ consisting of all pairs
$\gamma \to \gamma'$ for which $\gamma \Rightarrow \gamma'$ as defined by our
semantics, where $\gamma$ consists of a single statement $S_0$, and the active
statement being processed to get $\gamma'$ is a direct child of $S_0$.  Since
our semantics always substitutes siblings for siblings, it is clear that
$\Rightarrow_{\R}$ then coincides with our rewrite relation $\Rightarrow$.
Since the processing of every individual statement is always deterministic (up
to isomorphism of trees), $\R$ as just defined is clearly unequivocal.

We want to show that $\R$ is closed.  Thereto, we define residue
maps $r[\gamma,\gamma']$ as follows.

The case where $\gamma\to\gamma'$ is the processing of an apply- or
call-statement, is depicted in Figure~\ref{figrosen} (top).  The node being
processed is shown in black.  The subtemplates to the left and right are left
untouched.  Referring to the notation used in Figure~\ref{figapply}, the newly
substituted subtemplate $\gamma_{\mathrm{new}}$ is such that $\gamma_1 \dots
\gamma_k \normif \gamma_{\mathrm{new}}$ (for apply) or $\gamma_1 \normif
\gamma_{\mathrm{new}}$ (for call).  Indeed, since we apply $\normif$ at the
end of every processing step, $\gamma$ itself does not contain any active
if-statements.  We define $r = r[\gamma,\gamma']$ as follows:
\begin{itemize}
\item
For nodes $n$ in $\gamma_{\mathrm{left}}$ or $\gamma_{\mathrm{right}}$, we put
$r(n) := \{n'\}$, where $n'$ is the corresponding node in $\gamma'$.
\item
For the black node $b$, we put $r(b) := \emptyset$.
\end{itemize}
The main condition for closedness is clearly satisfied, because statements can
be processed independently.  Note that the black node has no children, let
alone active children, which allows us to put $r(b)=\emptyset$.  The condition
on $p$'s is also satisfied, because both $r[\phi_0,\psi_0]$ and
$r[\phi_1,\psi_1]$ will set $r(p)$ to $\{p'\}$.

The case where $\gamma\to\gamma'$ is the processing of a foreach-statement is
depicted in Figure~\ref{figrosen} (middle).  This case is analogous to the
previous one.  The only difference is that the black node now has descendants
($M$ in the figure).  Because the $\init$ function (Definition~\ref{definit})
always leaves descendants of a foreach node inactive, however, the nodes in
$M$ are inactive at this time, and we can
put $r(n) := \emptyset$ for all of them.

The case where $\gamma\to\gamma'$ is the processing of a val-statement is
depicted in Figure~\ref{figrosen} (bottom).  Since all
nodes in the update set are inactive by definition
(Definition~\ref{defupdateset}), we can again put $r(n) :=
\emptyset$ for all nodes in the update set.
The case of a tree-statement is similar; now the black node again
has descendants, but again these are all inactive (they are all
cons-statements).
\begin{figure}
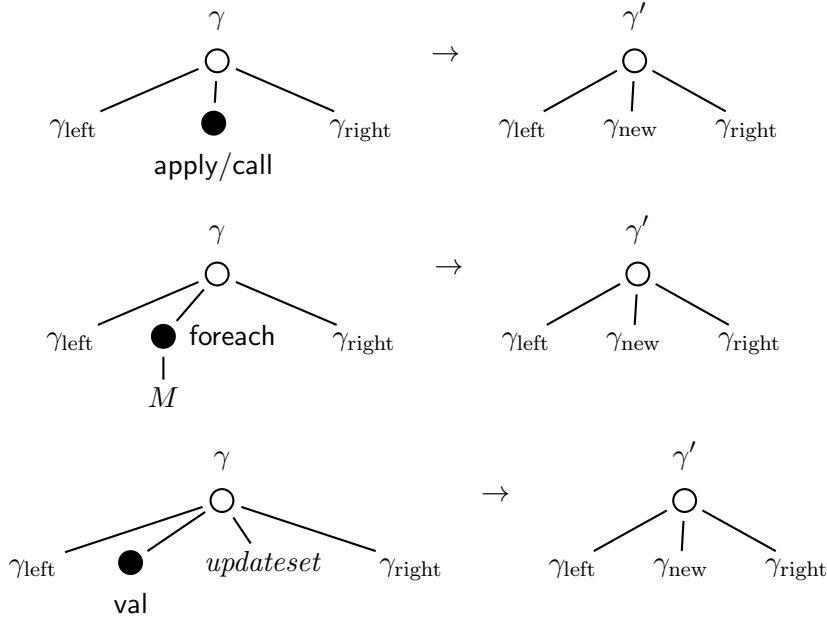

\begin{center}
\psset{tnpos=a}
\pstree{\TC~{$\gamma$}}
{
  \TR{$\gamma_{\mathrm{left}}$}
  \TC[fillstyle=solid]~[tnpos=b]{\textsf{apply/call}}
  \TR{$\gamma_{\mathrm{right}}$}
}
\quad
$\to$
\quad
\pstree{\TC~{$\gamma'$}}
{
  \TR{$\gamma_{\mathrm{left}}$}
  \TR{$\gamma_{\mathrm{new}}$}
  \TR{$\gamma_{\mathrm{right}}$}
}
\bigskip

\pstree{\TC~{$\gamma$}}
{
  \TR{$\gamma_{\mathrm{left}}$}
  \pstree{\TC[fillstyle=solid]~[tnpos=r]{\textsf{foreach}}}
  {
    \TR{$M$}
  }
  \TR{$\gamma_{\mathrm{right}}$}
}
\quad
$\to$
\quad
\pstree{\TC~{$\gamma'$}}
{
  \TR{$\gamma_{\mathrm{left}}$}
  \TR{$\gamma_{\mathrm{new}}$}
  \TR{$\gamma_{\mathrm{right}}$}
}
\bigskip

\pstree{\TC~{$\gamma$}}
{
  \TR{$\gamma_{\mathrm{left}}$}
  \TC[fillstyle=solid]~[tnpos=b]{\textsf{val}}
  \TR{\textit{updateset}}
  \TR{$\gamma_{\mathrm{right}}$}
}
\quad $\to$ \quad
\pstree{\TC~{$\gamma'$}}
{
  \TR{$\gamma_{\mathrm{left}}$}
  \TR{$\gamma_{\mathrm{new}}$}
  \TR{$\gamma_{\mathrm{right}}$}
}
\end{center}
\caption{Illustration to the proof of Theorem~\ref{confluence}.}
\label{figrosen}
\end{figure}
The case where $\gamma\to\gamma'$ is the processing of a
copy-statement, finally, is again analogous.
\end{proof}

\section{Computational completeness}

\newcommand{\allexp}{\texttt{//*}}
\newcommand{\firstchild}{\texttt{child::*[1]}}
\newcommand{\nextexp}{\texttt{following-sibling::*[1]}}
\newcommand{\flatree}{\mathit{flattree}}

As defined in Definition~\ref{deffinal},
an XSLT program $P$ expresses a partial
function from data trees to data forests, where the output forest is
represented by a tree by affixing a root node labeled $\ttdoc$ on top
(Definition~\ref{defmaketree}).  The output
is defined up to isomorphism only, and $P$ does not distinguish between
isomorphic inputs.  This leads us to the following definition:

\begin{definition}
A \emph{tree transformation}
is a partial function from data trees to data
trees with root labeled $\ttdoc$,
mapping isomorphic trees to isomorphic trees.
\end{definition}

Using the string representation of data trees defined in
Section~\ref{secdatatrees}, we further define:

\begin{definition}
A tree transformation $f$ is called \emph{computable} if the string function
$\tilde f\colon \tstring(\t) \mapsto \tstring(f(\t))$ is computable in the
classical sense.
\end{definition}

Up to now, we have assumed from our XPath abstraction $\X$ only the
availability of the expressions `\rootexp' and `\childexp'. For
our proof of the following
theorem, we need to assume the availability of a few more very simple
expressions, also present in real XPath:

\begin{itemize}

\item

$y$\texttt{/*},
for any tree variable $y$,
evaluates to the
root of the tree assigned to $y$.

\item

\allexp\ evaluates to the sequence of all nodes in the store (it does not
matter in which order).

\item

\firstchild\ evaluates to the first child of the context item (which should
be a node).

\item

\nextexp\ evaluates to the immediate right sibling of the context node, or
the empty sequence if the context node has no right siblings.

\item

Increment, decrement, and test on counters:
the constant expression
`\texttt{1}', and the expressions
`\texttt{$x$+1}', `\texttt{$x$-1}', and `\texttt{$x$=1}' for any value
variable $x$, which should consist of a single counter.  If
$x$ has the maximal counter value,  then
\texttt{$x$+1} need not be defined, and if 
$x$ has value 1, then
\texttt{$x$-1} need not be defined.  The test
\texttt{$x$=1} yields any nonempty sequence for true and the empty sequence
for false.

\item

\texttt{name()='a'}, for any $\tta \in \Sigma$, returning any nonempty
sequence if the label of the context node is \tta, and the empty sequence
otherwise.

\item

\texttt{()} evaluates to the empty sequence.

\end{itemize}

We establish:

\begin{theorem} \label{completeness}
Every computable tree transformation $f$ can be realised by a program.
\end{theorem}

\begin{proof}
We can naturally represent
any string $s$ over some finite alphabet as a flat data tree
over the same alphabet.  We denote this flat tree by $\flatree(s)$.
Its root is
labeled \ttdoc, and has $k$
children, where $k$ is the length of $s$, such that the labels of the children
spell out the string $s$.  There are no other nodes.

The proof now consists of three parts:
\begin{enumerate}
\item
Program the transformation $\t \mapsto \flatree(\tstring(\t))$.
\item
Show that every turing machine (working on strings)
can be simulated by some program working on the
$\flatree$ representation of strings.
\item
Program the transformation $\flatree(\tstring(\t)) \mapsto \t$.
\end{enumerate}
The theorem then follows by composing these three steps, where we simulate a
turing machine for $\tilde f$ in step~2. Note that the composition of three
programs can be written as a single program, using a temporary tree to pass
the intermediate results, and using modes to keep the rules from the different
programs separate.

The programs for steps 1 and 3 are shown in Figures \ref{figt2s} and
\ref{figs2t}. For simplicity, they are for an alphabet consisting of a single
letter \tta, but it is obvious how to generalise the programs.  The real XSLT
versions are given in the Appendix.  We point out that these programs are
actually 1.0 programs, so it is only for step~2 of the proof that we need
XSLT~2.0.

\begin{figure}
\begin{tabbing}
\quad\=\kill
\textsf{template \texttt{tree2string} match (\allexp)} \\
\aco \+\\
\textsf{cons \texttt{a} \aco\ \act} \\
\textsf{cons \texttt{lbrace} \aco\ \act} \\
\textsf{apply (\childexp)} \\
\textsf{cons \texttt{rbrace} \aco\ \act} \- \\
\act
\end{tabbing}
\caption{From $\t$ to $\flatree(\tstring(\t))$.}
\label{figt2s}
\end{figure}

\begin{figure}
\small
\begin{tabbing}
\quad\=\quad\=\quad\=\quad\=\quad\=\quad\=\kill
\textsf{template \texttt{doc} match (\rootexp)} \\
\aco \\
\>\textsf{apply (\firstchild)} \\
\act \\[\medskipamount]
\textsf{template \texttt{string2tree} match (\allexp)} \\
\aco \+ \\
\textsf{cons \texttt{a}} \\
\aco\ \textsf{apply (\nextexp) mode \texttt{dochildren}} \act \\
\textsf{val \texttt{counter} (\texttt{1})} \\
\textsf{call \texttt{searchnextsibling}} \-\\
\act \\[\medskipamount]
\textsf{template \texttt{dochildren} match (\allexp)
mode \texttt{dochildren}} \\
\aco \+ \\
\textsf{if \texttt{name()='lbrace'}} \\
\aco\ \textsf{apply (\nextexp) mode \texttt{dochildren}} \act \\
\textsf{else} \aco \+\\
\textsf{if \texttt{name()='a'}} \\
\aco\ \textsf{call \texttt{string2tree}} \act \\
\textsf{else} \aco\ \act \-\\
\act \-\\
\act \\[\medskipamount]
\textsf{template \texttt{searchnextsibling} match (\allexp) mode
\texttt{search}} \\
\aco \+ \\
\textsf{if \texttt{name()='lbrace'}} \aco\+\\
\textsf{val \texttt{counter} (\texttt{counter + 1})} \\
\textsf{apply (\nextexp) mode \texttt{search}} \-\\
\act \\
\textsf{else} \aco\+\\
\textsf{if \texttt{name()='a'}} \\
\aco\ \textsf{apply (\nextexp) mode \texttt{search}} \act\\
\textsf{else} \aco\+\\
\textsf{val \texttt{counter} (\texttt{counter - 1})} \\
\textsf{if \texttt{counter = 1}} \\
\aco\ \textsf{apply }\=\textsf{(\nextexp)} \\
\>\textsf{mode \texttt{dochildren}} \act \\
\textsf{else} \\
\aco\ \textsf{apply (\nextexp) mode \texttt{search}} \act \-\\
\act \-\\
\act \-\\
\act
\end{tabbing}
\caption{From $\flatree(\tstring(\t))$ to $\t$.}
\label{figs2t}
\end{figure}

For step~2, we can represent a configuration of a turing machine $A$ by two
temporary trees \texttt{left} and \texttt{right}.  At each step, variable
\texttt{right} holds (as a flat tree)
the content of the tape starting at the
head position and ending in the last tape cell;
variable \texttt{left} holds the reverse of the tape portion left of the head
position.  To keep track of the current state of the machine, we use value
variables $q$ for each state $q$ of $A$, such that at each step precisely one
of these is nonempty.  (This is why we need the $\X$-expression \texttt{()}.)
Changing the symbol under the head to an \texttt{a} amounts to assigning a new
content to \texttt{right} by putting in \textsf{cons \tta\ \aco\act},
followed by copies of the nodes in the current content of
\texttt{right}, where we skip the first one.
Moving the head a cell
to the right amounts to assigning a new content to \texttt{left} by putting
in a node labeled with the current symbol, followed by copies of the nodes in
the current content of \texttt{left}. We also assign a new content to
\texttt{right} in the now obvious way; if we were at the end of the
tape we add a new node labeled \texttt{blank}.
Moving the head a cell to the left is
simulated analogously. The only $\X$-expressions we need here are the ones we
have assumed to be available.

The simulation thus consists of repeatedly calling a big if-then-else that
tests for the transition to be performed, and performs that transition.  We
may assume $A$ is programmed in such a way that the final output is produced
starting from a designated state.  In this way we can build up the final
output string in a fresh temporary tree and pass it to step~3.
\end{proof}

\section{XSLT 1.0}

\newcommand{\G}{\mathcal{G}}

In this section we will show that every XSLT~1.0 program can be implemented in
exponential time, in sharp contrast to the computational completeness result
of the previous section.

A fundamental difference between XSLT 1.0 and 2.0 is that in 1.0,
$\X$-expressions are ``input-only'', defined as follows.

\begin{definition} \begin{enumerate} \item Let $C=(\S,\E,(z,i,k))$ be a
context. Let the input tree in $\S$ be $\t$. Then we call $C$
\emph{input-only} if every value appearing in $\E$ is already a value over the
store $\{(\Input,\t)\}$, and also $(z,i,k)$ is like that.
\item
By $\hat C$, we mean
the context $(\{(\Input,\t)\},\E,(z,i,k))$.  So, $\hat C$ equals $C$ where we
have removed all temporary trees.
\item
Now an $\X$-expression $e$ is called input-only if for any input-only context
$C$ for which $\eval(e,C)$ is defined, we have $\eval(e,C) = \eval(e,\hat C)$,
and this must be a value over $C$'s input tree only.
\end{enumerate} \end{definition}

In other words, input-only expressions are oblivious to the temporary trees in
the store; they only see the input tree.

We further define:

\begin{definition} An input-only $\X$-expression $e$ is called
\emph{polynomial} if for each input-only context $C$, the computation of
$\eval(e,C)$ can be done in time polynomial in the size of $C$'s input tree.
\end{definition}

We now define:

\begin{definition}
A program is called 1.0 if it only uses input-only, polynomial
$\X$-expressions.
\end{definition}

Essentially, 1.0 programs cannot do anything with temporary trees except copy
them using \textsf{tcopy} statements.  We note that real XPath~1.0 expressions
are indeed input-only and polynomial; actually, real XPath~1.0 is much more
restricted than that, but for our purpose we do not need to assume anything
more.

In order to establish an exponential upper bound on the time-complexity of 1.0
programs, we cannot use an explicit representation of the output tree.
Indeed, 1.0 programs can produce result trees of size \emph{doubly}
exponential in the size of the input tree.  For example, using subsets of
input nodes, ordered lexicographically, as depth counters, we can produce a
full binary tree of depth $2^n$ from an input tree with $n$ nodes.  Obviously
a doubly exponentially long output could never be computed in singly
exponential time.

\begin{figure}
\begin{center}
\includegraphics{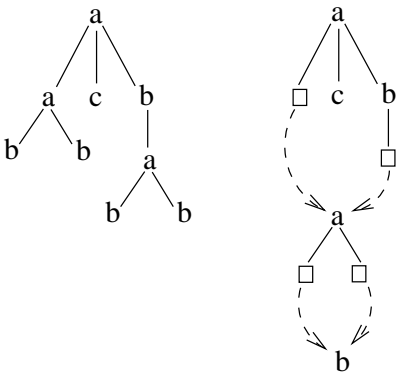}
\end{center}
\caption{Left, a data tree, and right, a DAG representation of it.}
\label{figdag}
\end{figure}

We therefore use a \emph{DAG representation} of trees: an old and well-known
trick \cite{linearunification} that is also used in tree transduction
\cite{maneth_macromplexity}, and that has recently found new
applications in XML \cite{bgk_xmlcompressed}.
Formally, a DAG representation is a collection $\G$ of trees, where trees in
$\G$ can have special leafs which are not labeled, and from which a pointer
departs to the root of another tree in $\G$. On condition that the resulting
pointer graph is acyclic, starting from a designated ``root tree'' in $\G$ we
can naturally obtain a tree by unfolding along the pointers.  An illustration
is shown in Figure~\ref{figdag}.

We establish:

\begin{theorem} \label{complexity}
Let $P$ be an 1.0 program. Then the following problem is solvable in
exponential, i.e., $2^{n^{O(1)}}$ time:
\begin{description}
\item[Input:] a data tree $\t$
\item[Output:] a DAG representation of the final result tree of applying $P$
to $\t$, or a message signaling non-termination if $P$ does not terminate on
$\t$.
\end{description}
\end{theorem}

\begin{proof}
We will generate a DAG representation $\G$ by applying modified versions of
the semantic rules from Section~\ref{secsemantics}.  We initialise $\G$ with
all the subtrees of $\t$.  These trees have no pointers.  Each tree
that will be added to $\G$ will be a configuration, which still has to be
developed further into a final data tree with pointers, using the same
modified rules.  Because we will have to point to the newly added
configurations later, we identify each added configuration by a pair
$(\mathit{name},C)$ where $\mathit{name}$ is the name of a template rule in
$P$ and $C$ is a context.  In the description below, whenever we say that we
``add'' a configuration to $\G$, identified by some pair $(\mathit{name},C)$,
we really mean that we add it unless a configuration identified by that same
pair already exists in $\G$.

The modifications are now the following.

\begin{enumerate}

\item

When executing an apply-statement, we do not directly insert copies of the
templates belonging to the rules that must be applied (the $\gamma_i$'s in
Figure~\ref{figapply}). Rather, we add, for $i=1,\dots,k$, the configuration
$\gamma_i'$ to $\G$, where $\gamma_i \normif \gamma_i'$.  We identify
$\gamma_i'$ by the pair $(\mathit{name}_i,C_i)$, with $\mathit{name}_i$ the
name of the rule $\gamma_i$ comes from, and $C_i = (\S,\E,(\n_i,i,k))$ using
the notation of Figure~\ref{figapply}.  Moreover, in place of the
apply-statement we insert a sequence of
$k$ pointer nodes pointing to $(\mathit{name}_1,C_1)$, \dots,
$(\mathit{name}_k,C_k)$, respectively.

\item

When executing a call-statement \textsf{call $\mathit{name}$} under context
$C$, we again do not insert $\gamma_1$ (compare Figure~\ref{figapply}), but
add the configuration $\gamma_1'$ to $\G$, where $\gamma_1 \normif \gamma_1'$,
and identify it by the pair $(\mathit{name},C)$. We then replace the
statement by a pointer node pointing to that pair.

\item

By making template rules from the bodies of all foreach-statements in $P$, we
may assume without loss of generality that the body of every foreach-statement
is a single call-statement.  A foreach-statement is then processed analogously
to apply- and call-statements.

\item

As we did with foreach-statements, we may assume that the body of each
tree-statement is a single call-statement.  When executing a tree-statement,
we may assume that the call-statement has already been turned into a pointer
to some pair $(\mathit{name}_0,C_0)$. We then assign that pair directly to $y$
in the new context $C'$ (compare Figure~\ref{figvar}); we no longer apply
$\maketree$.

So, in the modified kind of store we use, we assign name--context pairs,
rather than fully specified temporary trees, to tree variables.

\item

Correspondingly, when executing a statement \textsf{tcopy $y$}, we now
directly turn it into a pointer to the pair assigned to $y$.

\item

Finally, when executing a vcopy-statement, we do not insert the whole
forest generated by $(\n_1,\dots,\n_k)$
in the configuration (compare
Figure~\ref{figcopy}), but merely insert a sequence of
$k$ pointers to the input subtrees rooted at $\n_1$, \dots, $\n_k$,
respectively.

\end{enumerate}

We initiate the generation of $\G$ by starting with the initial configuration
as always.  Processing that configuration will add the first tree to $\G$,
which serves as the root tree of the DAG representation.  When all trees in
$\G$ have been fully developed into data trees with pointer nodes, the
algorithm terminates.  In case $P$ does not terminate on $\t$, however, that
will never happen, and we need a way to detect nontermination.

Thereto, recall that every context consists of an environment $\E$ and a
context triple $c$ on the one hand, and a store $\S$ on the other hand.  Since
all $\X$-expressions used are input-only, and thus oblivious to the store-part
of a context (except for the input tree, which does not change), we are in an
infinite loop from the moment that there is a cycle in $\G$'s pointer graph
\emph{where we ignore the store-part of the contexts}.  More precisely, this
happens when from a pointer node in a tree identified by $(\mathit{name},C_1)$
we can follow pointers and reach a pointer to a pair $(\mathit{name},C_2)$
with the same $\mathit{name}$ and where $C_1$ and $C_2$ are equal in their
$(\E,c)$-parts.  As soon as we detect such a cycle, we terminate the algorithm
and signal nontermination.  Note that thus the algorithm always terminates.
Indeed, since only input-only $\X$-expressions are used, all contexts that
appear in the computation are input-only, and there are only a finite number
of possible $(\E,c)$-parts of input-only configuration over a fixed input
tree.

Let us analyse the complexity of this algorithm.  Since all $\X$-expressions
used are polynomial, there is a natural number $K$ such that each value that
appears in a context is at most $n^K$ long, where $n$ equals the number of
nodes in $\t$.  Each element of such a length-$n^K$ sequence is a node or a
counter over $\t$, so there are at most $(2n)^{n^K}$ different values.  There
are a constant $c_1$ number of different value variables in $P$, so there are
at most $((2n)^{n^K})^{c_1}$ different environments.  Likewise, the number of
different context triples is $(2n)^3$, so, ignoring the stores, there are in
total at most $(2n)^3 \cdot (2n)^{c_1 n^K} \leq 2^{n^{K'}}$ different
contexts, for some natural number $K' \geq K$.
With a constant $c_2$ number of different template names in $P$, we get a
maximal number of $c_2 2^{n^{K'}}$ different configurations that can be added
to $\G$ before the algorithm will surely terminate.

It remains to see how long it takes to fully rewrite each of those
configurations into a data tree with pointers.  A configuration initially
consists of at most a constant $c_3$ number of statements.  The evaluation of
$\X$-expressions, which are polynomial, takes at most $c_3 n^K$ time in total.
Processing an apply- or a foreach-statement takes at most $c_3n^K$
modifications to the configuration and to $\G$; for the other statements this
takes at most $c_3$ such operations.  Each such operation, however, involves
the handling of contexts, whose stores can become quite large if treated
naively.  Indeed, tree-statements assign a context to a tree variable,
yielding a new context which may then again be assigned to a tree variable,
and so on.  To keep this under control, we do not copy the contexts literally,
but number them consecutively in the order they are introduced in $\G$.  A map
data structure keeps track of this numbering.  The stores then consist of an
at most constant $c_4$ number of assignments of pairs (name, context number)
to tree variables.  As there are at most $2^{n^{K'}}$ different contexts, each
number is at most $n^{K'}$ bits long.  Looking up whether a given context is
already in $\G$, and if so, finding its number, takes $O(\log 2^{n^{K'}}) =
O(n^{K'})$ time using a suitable map data structure.

We conclude that the processing of $\G$ takes a total time of $c_2 2^{n^{K'}}
\cdot O(n^{K'}) = 2^{n^{O(1)}}$, as had to be proven.
\end{proof}

A legitimate question is whether the complexity bound given by
Theorem~\ref{complexity} can still be improved.  In this respect we can show
that, even within the limits of real XSLT~1.0, any linear-space turing machine
can be simulated by a 1.0 program.  Note that some PSPACE-complete problems,
such as QBF-SAT \cite{papadimitriou}, are solvable in linear space. This shows
that the time complexity upper bound of Theorem~\ref{complexity} cannot be
improved without showing that PSPACE is properly included in EXPTIME (a famous
open problem).

The simulation gets as input a flat tree representing an input string, and
uses the $n$ child nodes to simulate the $n$ tape cells.  For each letter
\tta\ of the tape alphabet, a value variable $\mathit{cell}_{\text{\tta}}$
holds the nodes representing the tape cells that have an \tta.  A value
variable $\it head$ holds the node representing the cell seen by the machine's
head.  The machine's state is kept by additional value variables ${\it
state}_q$ for each state $q$, such that ${\it state}_q$ is nonempty iff the
machine is in state $q$.  Writing a letter in a cell, moving the head left or
right, or changing state, are accomplished by easy updates on the
value-variables, which can be expressed by real XPath~1.0 expressions.
Choosing the right transition is done by a big if-then-else statement.
Successive transitions are performed by recursively applying the simulating
template rule until a halting state is reached.

\begin{remark}
A final remark is that our results imply that XSLT~1.0 is not closed under
composition.  Indeed, building up a tree of
doubly exponential size (as we already remarked is possible in XSLT~1.0),
followed by the building up of a tree of exponential
size, amounts to building up a tree of triply exponential size.  If that would
be possible by a single program, then a DAG representation of a triply
exponentially large tree would be computable
in singly exponential time.  It is well known, however, that a DAG
representation cannot be more than singly
exponentially smaller than the tree it
represents.  Closure under composition is another sharp contrast
between XSLT~1.0 and 2.0, as the latter is indeed
closed under composition as already noted in the proof of
Theorem~\ref{completeness}.
\end{remark}

\section{Conclusions}

W3C recommendations such as the XSLT specifications are no Holy scriptures.
Theoretical scrutinising of W3C work, which is what we have done here, can
help in better understanding the possibilities and limitations of various
newly proposed programming languages related to the Web, eventually leading to
better proposals.

A formalisation of the full XSLT~2.0 language, with all the dirty details both
concerning the language itself as concerning the XPath~2.0 data model, is
probably something that should be done.  We believe our work gives a clear
direction how this could be done.

Note also that XSLT contains a lot of redundancies.  For example,
foreach-statements are eliminable, as are call-statements, and the match
attribute of template rules.  A formalisation such as ours can provide a
rigorous foundation to prove such redundancies, or to prove correct various
processing strategies or optimisation techniques XSLT implementations may use.

A formal tree transformation model denoted by TL, in part inspired by XSLT,
but still omitting many of its features, has already been studied by Maneth
and his collaborators \cite{frank_xsltformal,maneth_pods}.  The TL model can
be compiled into the earlier formalism of ``macro tree transducers''
\cite{mtt,ps_macroforest}.  It is certainly an interesting topic for further
research to similarly translate our XSLT formalisation (even partially) into
macro tree transducers, so that techniques already developed for these
transducers can be applied.  For example, under regular expression types
\cite{xduce} (known much earlier under the name of ``recognisable tree
languages''), exact automated typechecking is possible for compositions of
macro tree transducers, using the method of ``inverse type inference''
\cite{msv_typecheckxml}.  This method has various other applications, such as
deciding termination on all possible inputs \cite{maneth_pods}.  Being able to
apply this method to our XSLT~1.0 formalism would improve the analysis
techniques of Dong and Bailey \cite{bailey_xslt}, which are not complete.

\section*{Acknowledgment}

We are indebted to Frank Neven for his initial participation in this research.

\bibliographystyle{plain}

\clearpage
\appendix

\section{Real XSLT programs}
	\subsection{Figure~\ref{figt2s} in real XSLT}

	\begin{alltt}


	<xsl:transform 
	  xmlns:xsl="http://www.w3.org/1999/XSL/Transform" 
	  version="1.0">
	
	  <xsl:template name="tree2string" match="//*">
	    <a/>
	    <lbrace/>
	    <xsl:apply-templates select="child::*"/>
	    <rbrace/>
	  </xsl:template>
	
	</xsl:transform>

	\end{alltt}

	\subsection{Figure~\ref{figs2t} in real XSLT}

\begin{alltt}

<xsl:transform 
  xmlns:xsl="http://www.w3.org/1999/XSL/Transform"
  version="1.0">

<xsl:template match="/doc">
  <xsl:apply-templates select="child::*[1]"/>
</xsl:template>

<xsl:template name="string2tree" match="/doc//*">
  <a>
    <xsl:apply-templates select="following-sibling::*[1]" mode="dochildren"/>
  </a>
  <xsl:call-template name="searchnextsibling">
    <xsl:with-param name="counter" select="1"/>
  </xsl:call-template>
</xsl:template>

<xsl:template match="//*" mode="dochildren">
  <xsl:if test="name()='lbrace'">
    <xsl:apply-templates select="following-sibling::*[1]" mode="dochildren"/>
  </xsl:if>
  <xsl:if test="name()='a'">
    <xsl:call-template name="string2tree"/>
  </xsl:if>
</xsl:template>

<xsl:template name="searchnextsibling" match="//*" mode="search">
  <xsl:param name="counter"/>
  <xsl:if test="name()='lbrace'">
    <xsl:apply-templates select="following-sibling::*[1]" mode="search">
      <xsl:with-param name="counter" select="$counter + 1"/>
    </xsl:apply-templates>
  </xsl:if>
  <xsl:if test="name()='a'">
    <xsl:apply-templates select="following-sibling::*[1]" mode="search">
      <xsl:with-param name="counter" select="$counter"/>
    </xsl:apply-templates>
  </xsl:if>
  <xsl:if test="name()='rbrace'">
    <xsl:if test="$counter=2">
      <xsl:apply-templates select="following-sibling::*[1]" mode="dochildren"/>
    </xsl:if>
    <xsl:if test="$counter>2">
      <xsl:apply-templates select="following-sibling::*[1]" mode="search">
        <xsl:with-param name="counter" select="$counter - 1"/>
      </xsl:apply-templates>
    </xsl:if>
  </xsl:if>
</xsl:template>

</xsl:transform>

\end{alltt}

\end{document}